\documentstyle[amsfonts,aps, pra]{revtex}

\begin{document}
\title{Complex Ginzburg--Landau equation for self--oscillatory systems under
amplitude modulated forcing in a 1:1 resonance}
\author{Germ\'{a}n J. de Valc\'{a}rcel}
\address{Departament d'\`{O}ptica, Universitat de Val\`{e}ncia,\\
Dr. Moliner 50, 46100-Burjassot, Spain.\\
email: german.valcarcel@uv.es}
\maketitle

\begin{abstract}
The dynamics of self-oscillatory extended systems, resonantly forced at a
frequency close to that of the natural oscillations (1:1 resonance), is
shown to be universally described by a complex Ginzburg--Landau equation
containing an inhomogeneous term. The case of amplitude modulated forcing is
considered, which generalizes previous studies. Application to the
two-frequency forcing in a 1:1 resonance is considered as an example.
\end{abstract}

\section{Introduction}

The periodic forcing of spatially extended self-oscillatory systems is a
classical method to excite the formation of spatial patterns in such
systems. This kind of forcing admits a universal description when the system
is operated near the oscillation threshold, and forcing acts on a $n:m$
resonance, defined by the relation $\omega _{e}=\left( n/m\right) \left(
\omega _{0}+\nu \right) $ between the external forcing frequency $\omega _{e}
$ and the natural frequency of oscillations $\omega _{0}$, where $n/m$ is an
irreducible integer fraction and $\nu $ is a small mistuning. In such case
the slowly varying complex amplitude of the oscillations $u$ verifies the
following generalized complex Ginzburg-Landau equation (CGLE) \cite{coullet} 
\begin{equation}
\partial _{t}u=a_{1}u+a_{2}\nabla ^{2}u+a_{3}\left| u\right| ^{2}u+a_{4}%
\bar{u}^{n-1},  \label{cglcoullet}
\end{equation}
where $\bar{u}$ stands for the complex conjugate of $u$, $a_{i=1,2,3}$ are
complex coefficients and $a_{4}$ is proportional to the $m$th power of the
forcing amplitude. Eq. (\ref{cglcoullet}) is valid in principle for
perfectly periodic forcings and has been introduced making use of symmetry
arguments \cite{coullet}. That elegant reasoning is based on the existence
of a discrete time invariance when a perfectly periodic forcing is applied
to a system. However if, {\it e.g.}, the amplitude of forcing is modulated
in time in an arbitrary way all temporal symmetries are broken and Eq. (\ref
{cglcoullet}) is not rigorously justified. In this work we prove that Eq. (%
\ref{cglcoullet}) does provide a universal description of the close to
threshold dynamics of self-oscillatory extended systems forced in a 1:1
resonance ($n=1$ and $m=1$ in Eq. (\ref{cglcoullet})). The analysis is based
on the technique of multiple scales and generalizes the concept of resonant
forcing as it considers (almost) periodic forcings that are nonuniform
across the system and/or modulated in time. Our derivation hence allows a
rigorous simplified study of both noisy forcings as well as multi-frequency
forcings within a 1:1 resonance. Generalizations to other resonances ($n\neq
1$ or $m\neq 1$), although cumbersome, can be made straightforwardly
following the lines of the present derivation. 

\section{Model}

We consider a generic two-dimensional system described by $N$ real dynamical
variables $\left\{ U_{i}\left( {\bf r},t\right) \right\} _{i=1}^{N}$ whose
time evolution is governed by the following set of real equations written in
vector form, 
\begin{equation}
\partial _{t}{\bf U}\left( {\bf r},t\right) ={\bf f}\left( \mu ,\alpha ;{\bf %
U},\nabla ^{2}{\bf U}\right) ,  \label{model}
\end{equation}
where ${\bf f}$ is a sufficiently differentiable function of its arguments.
We assume that the dependence of ${\bf f}$ on spatial derivatives of ${\bf U}
$ is through its Laplacian $\nabla ^{2}=\partial _{x}^{2}+\partial _{y}^{2}$%
, ${\bf r}=\left( x,y\right) $. This is the simplest dependence on
derivatives in rotationally invariant systems at the time it corresponds to\
physical systems of most relevance like, {\it e.g.}, reaction-diffusion and
nonlinear optical systems. $\mu $ is the bifurcation parameter and $\alpha
\left( {\bf r},t\right) $ is the forcing parameter, which is allowed to vary
on time and space. Physically $\alpha $ may represent either an independent
parameter, or the modulated part of any other parameter.

We assume that in the absence of forcing ($\alpha =0$) Eq. (\ref{model})
supports a steady, spatially homogeneous state ${\bf U}={\bf U}_{s}\left(
\mu \right) $ $\left( \partial _{t}{\bf U}_{s}=\partial _{x}{\bf U}%
_{s}=\partial _{y}{\bf U}_{s}=0\right) $, which looses stability at $\mu
=\mu _{0}$ giving rise to a self-oscillatory, spatially homogeneous state.
In other words, we assume that the reference state ${\bf U}_{s}$ suffers a
homogeneous Hopf bifurcation at $\mu =\mu _{0}$. We wish to study the small
amplitude solutions that form in the system close to the bifurcation when
the arbitrary parameter $\alpha $ is modulated in time with a frequency
close to that of the free oscillations.

For the sake of convenience we introduce a new vector 
\begin{equation}
{\bf u}\left( {\bf r},t\right) ={\bf U}\left( {\bf r},t\right) -{\bf U}_{s},
\label{u}
\end{equation}
which measures the deviation of the system from the reference state, in
terms of which we rewrite Eq. (\ref{model}) as a Taylor series, 
\begin{eqnarray}
\partial _{t}{\bf u} &=&{\bf F}\left( \mu ,\alpha \right) +{\cal J}\left(
\mu ,\alpha \right) \cdot {\bf u}+{\cal D}\left( \mu ,\alpha \right) \cdot
\nabla ^{2}{\bf u}  \nonumber \\
&&+{\bf K}\left( \mu ,\alpha ;{\bf u},{\bf u}\right) +{\bf L}\left( \mu
,\alpha ;{\bf u},{\bf u},{\bf u}\right) +{\rm h.o.t.},  \label{modelu}
\end{eqnarray}
where ${\rm h.o.t.}$ denotes terms of higher order than $3$ in ${\bf u}$ or
than $1$ in $\nabla ^{2}{\bf u}$. These ${\rm h.o.t.}$ have no influence
near the bifurcation $\left( \mu \simeq \mu _{0}\right) $ since, as we show
below, they are ${\cal O}\left( \left| \mu -\mu _{0}\right| ^{2}\right) $ or
smaller and only terms up to ${\cal O}\left( \left| \mu -\mu _{0}\right|
^{3/2}\right) $ contribute to the leading order dynamics of the system
whenever the bifurcation is supercritical, which is the case we assume.

The different elements of the expansion (\ref{modelu}) are defined as 
\begin{eqnarray*}
{\bf F}\left( \mu ,\alpha \right)  &=&{\bf f}_{s}, \\
{\cal J}_{ij}\left( \mu ,\alpha \right)  &=&\left[ \partial f_{i}/\partial
U_{j}\right] _{s},{\cal D}_{ij}\left( \mu ,\alpha \right) =\left[ \partial
f_{i}/\partial \nabla ^{2}U_{j}\right] _{s}, \\
{\bf K}\left( \mu ,\alpha ;{\bf a},{\bf b}\right)  &=&%
{\textstyle{1 \over 2!}}%
\mathop{\textstyle\sum}%
\nolimits_{i,j=1}^{N}\left[ \partial ^{2}{\bf f}/\partial U_{i}\partial U_{j}%
\right] _{s}a_{i}b_{j}, \\
{\bf L}\left( \mu ,\alpha ;{\bf a},{\bf b},{\bf c}\right)  &=&%
{\textstyle{1 \over 3!}}%
\mathop{\textstyle\sum}%
\nolimits_{i,j,k=1}^{N}\left[ \partial ^{3}{\bf f}/\partial U_{i}\partial
U_{j}\partial U_{k}\right] _{s}a_{i}b_{j}c_{k},
\end{eqnarray*}
where ${\bf a},{\bf b},{\bf c}$ are arbitrary vectors and the subscript $s$
denotes ${\bf U}={\bf U}_{s}\left( \mu \right) $. Vector ${\bf F}\left( \mu
,\alpha \right) $ is subjected to the condition 
\begin{equation}
{\bf F}\left( \mu ,0\right) =0,  \label{Fmu0}
\end{equation}
since in the absence of forcing ($\alpha =0$) the reference state ${\bf u}=0$
is a steady state of Eq. (\ref{modelu}) by hypothesis. ${\cal J}$ and ${\cal %
D}$ are matrices, and vector ${\bf K}$ (${\bf L}$) is a symmetric and
bilinear (symmetric and trilinear) function of its two (three) last
arguments.

\section{The Hopf bifurcation}

In the absence of forcing the stability of the reference state against small
perturbations $\delta {\bf u}$ is governed by the following equation 
\begin{equation}
\partial _{t}\delta {\bf u}={\cal J}\left( \mu ,0\right) \cdot \delta {\bf u}%
+{\cal D}\left( \mu ,0\right) \cdot \nabla ^{2}\delta {\bf u,}  \label{eqlin}
\end{equation}
obtained upon linearizing Eq. (\ref{modelu}) for $\alpha =0$ with respect to 
$\delta {\bf u}$. The general solution to Eq. (\ref{eqlin}) is a
superposition of plane waves of the form
\begin{equation}
\delta {\bf u}\left( {\bf r},t\right) =%
\mathop{\textstyle\sum}%
\nolimits_{j}{\bf w}_{j}\exp \left( \Lambda _{j}t\right) \exp \left( i{\bf k}%
_{j}{\bf \cdot r}\right) ,
\end{equation}
with 
\begin{eqnarray}
\Lambda _{j}{\bf w}_{j} &=&{\cal M}\left( \mu ,k_{j}^{2}\right) \cdot {\bf w}%
_{j},  \label{eigensystem} \\
{\cal M}\left( \mu ,k^{2}\right)  &=&{\cal J}\left( \mu ,0\right) -k^{2}%
{\cal D}\left( \mu ,0\right) ,  \label{matrixM}
\end{eqnarray}
($k^{2}={\bf k\cdot k}$) hence eigenvalues and eigenvectors of matrix ${\cal %
M}$ depend on ${\bf k}$ only through $k^{2}$ as ${\cal M}$ does. As we are
assuming that the reference state looses stability at $\mu =\mu _{0}$ via a
homogeneous Hopf bifurcation in the absence of forcing, matrix ${\cal M}%
\left( \mu ,k^{2}\right) $ must have a pair of complex-conjugate eigenvalues 
$\left\{ \Lambda _{1},\Lambda _{2}\right\} =\left\{ \lambda \left( \mu
,k^{2}\right) ,\bar{\lambda}\left( \mu ,k^{2}\right) \right\} $ (the overbar
denotes complex conjugation) governing the instability, {\it i.e.}:

\begin{enumerate}
\item[(i)]  Close to the bifurcation $%
\mathop{\rm Re}%
\Lambda _{i\geq 3}<0$ whilst $%
\mathop{\rm Re}%
\lambda $ can become positive for some $k$'s,

\item[(ii)]  At the bifurcation $%
\mathop{\rm Re}%
\lambda $ is maximum and null at $k=0$ (the perturbation with largest growth
rate is spatially homogeneous): 
\begin{equation}
\mathop{\rm Re}%
\lambda _{0}=0,\left( \partial _{k}%
\mathop{\rm Re}%
\lambda \right) _{0}=0,\left( \partial _{k}^{2}%
\mathop{\rm Re}%
\lambda \right) _{0}<0.  \label{Relambda0}
\end{equation}
where, here and in the following,
\end{enumerate}

\begin{center}
$\fbox{ \ subscript $0$ affecting functions denotes $\left\{ \mu =\mu
_{0},\alpha =0,k=0\right\} 
\begin{array}{c}
.
\end{array}
$ \ }$\\[0pt]
\end{center}

\begin{enumerate}
\item[(iii)]  The instability is oscillatory, {\it i.e.}, 
\begin{equation}
\mathop{\rm Im}%
\lambda _{0}=\omega _{0}\neq 0.  \label{Imlambda0}
\end{equation}
\end{enumerate}

Finally, the preceding properties imply that:

\begin{enumerate}
\item[(iv)]  All eigenvalues of ${\cal M}_{0}={\cal J}_{0}$, see Eq. (\ref
{matrixM}), have negative real part but $\left\{ \lambda _{0},\overline{%
\lambda }_{0}\right\} =\left\{ i\omega _{0},-i\omega _{0}\right\} $.
\end{enumerate}

For the sake of later use we introduce the right and left eigenvectors of $%
{\cal J}_{0}$ associated with eigenvalues $\left\{ i\omega _{0},-i\omega
_{0}\right\} $, 
\begin{equation}
\begin{array}{cc}
{\cal J}_{0}\cdot {\bf h}=i\omega _{0}{\bf h,} & {\cal J}_{0}\cdot \overline{%
{\bf h}}=-i\omega _{0}\overline{{\bf h}}, \\ 
{\bf h}^{\dagger }\cdot {\cal J}_{0}=i\omega _{0}{\bf h}^{\dagger }, & {\bf %
\,}\overline{{\bf h}^{\dagger }}\cdot {\cal J}_{0}=-i\omega _{0}\overline{%
{\bf h}^{\dagger }},
\end{array}
\label{r}
\end{equation}
where the short-hand notation ${\bf h}={\bf w}_{1}\left( \mu =\mu
_{0},k^{2}=0\right) ,\overline{{\bf h}}={\bf w}_{2}\left( \mu =\mu
_{0},k^{2}=0\right) $ has been introduced. These vectors verify the
following orthonormality relations: 
\[
{\bf h}^{\dagger }\cdot \overline{{\bf h}}=0,{\bf h}^{\dagger }\cdot {\bf h}%
=1\text{.} 
\]

\section{Scales}

We are interested in determining the small amplitude solutions that from in
the system close to the Hopf bifurcation, which we define by 
\begin{equation}
\mu =\mu _{0}+\varepsilon ^{2}\mu _{2},  \label{mu}
\end{equation}
where $\varepsilon $ is a smallness parameter $\left( 0<\varepsilon \ll
1\right) $. The study is based on the widely used technique of multiple
scales. These spatial and temporal scales appear naturally close to the
bifurcation and are those on which the asymptotic dynamics of the system
evolves. As is well known, in a homogeneous Hopf bifurcation these slow
scales are given by 
\begin{equation}
T=\varepsilon ^{2}t,X=\varepsilon x,Y=\varepsilon y.  \label{scales}
\end{equation}
These scales follow from the behaviour of $\lambda $ close to the
bifurcation, Eq. (\ref{mu}), for values of $k$ close to the most unstable
mode $k=0$:
\begin{equation}
\lambda \left( \mu _{0}+\varepsilon ^{2}\mu _{2},k^{2}\right) =\lambda
_{0}+\left( \partial _{\mu }\lambda \right) _{0}\varepsilon ^{2}\mu _{2}+%
\frac{1}{2}\left( \partial _{k}^{2}\lambda \right) _{0}k^{2}+\max \left\{ 
{\cal O}\left( \varepsilon ^{4}\right) ,{\cal O}\left( k^{2}\right) \right\}
,  \label{lambda}
\end{equation}
where the term $\left( \partial _{k}\lambda \right) _{0}k$ has not been
included since $\left( \partial _{k}\lambda \right) _{0}=0$ as $\lambda $ is
an even function of $k$. From this equation we obtain, making use of Eq. (%
\ref{Relambda0}), 
\[
\mathop{\rm Re}%
\lambda \left( \mu _{0}+\varepsilon ^{2}\mu _{2},k^{2}\right) =\left(
\partial _{\mu }%
\mathop{\rm Re}%
\lambda \right) _{0}\varepsilon ^{2}\mu _{2}-\frac{1}{2}\left| \partial
_{k}^{2}%
\mathop{\rm Re}%
\lambda \right| _{0}k^{2}+\max \left\{ {\cal O}\left( \varepsilon
^{4}\right) ,{\cal O}\left( k^{2}\right) \right\} ,
\]
what indicates that the only modes which can experience linear growth verify 
$k={\cal O}\left( \varepsilon \right) $, hence the asymptotic dynamics of
the system exhibits spatial variations on a scale $x,y\sim k^{-1}\sim
\varepsilon ^{-1}$and the slow spatial scales $\left( X=\varepsilon
x,Y=\varepsilon y\right) $ follow. Thus, putting $k=\varepsilon k_{1}$, 
\[
\mathop{\rm Re}%
\lambda \left( \mu _{0}+\varepsilon ^{2}\mu _{2},\varepsilon
^{2}k_{1}^{2}\right) =\varepsilon ^{2}\left[ \left( \partial _{\mu }%
\mathop{\rm Re}%
\lambda \right) _{0}\mu _{2}-\frac{1}{2}\left| \partial _{k}^{2}%
\mathop{\rm Re}%
\lambda \right| _{0}k_{1}^{2}\right] +{\cal O}\left( \varepsilon ^{4}\right)
,
\]
which shows that the growth of perturbations occurs on a scale $t\sim \left( 
\mathop{\rm Re}%
\lambda \right) ^{-1}\sim \varepsilon ^{-2}$ and the slow timescale $%
T=\varepsilon ^{2}t$ follows. On the other hand, making use of Eq. (\ref
{Imlambda0}) and putting $k=\varepsilon k_{1}$ again, Eq. (\ref{lambda})
yields 
\[
\mathop{\rm Im}%
\lambda \left( \mu _{0}+\varepsilon ^{2}\mu _{2},\varepsilon
^{2}k_{1}^{2}\right) =\omega _{0}+\varepsilon ^{2}\left[ \left( \partial
_{\mu }%
\mathop{\rm Im}%
\lambda \right) _{0}\mu _{2}+\frac{1}{2}\left( \partial _{k}^{2}%
\mathop{\rm Im}%
\lambda \right) _{0}k_{1}^{2}\right] +{\cal O}\left( \varepsilon ^{4}\right)
,
\]
whose first term, $\omega _{0}={\cal O}\left( \varepsilon ^{0}\right) $,
indicates that the original timescale $t$ must be retained, whilst the rest
of terms do not introduce other relevant timescales.

As for the external forcing we assume that its form is consistent with the
previous scales. In particular we assume that it is weak and of the form 
\begin{eqnarray}
\alpha \left( {\bf r},t\right) &=&\varepsilon ^{3}\alpha _{3}\left(
X,Y,T,t\right) ,  \label{alpha} \\
\alpha _{3}\left( X,Y,T,t\right) &=&\sum\nolimits_{p=1}^{\infty }\left[
\alpha _{3,p}\left( X,Y,T\right) \exp \left( ip\omega _{0}t\right) +\bar{%
\alpha}_{3,p}\left( X,Y,T\right) \exp \left( -ip\omega _{0}t\right) \right] ,
\label{alpha3}
\end{eqnarray}
Thus we are considering a resonant forcing with slowly varying amplitude.
Notice that, according to Eqs. (\ref{alpha},\ref{alpha3}), $\alpha \left( 
{\bf r},t+\frac{2\pi }{\omega _{0}}\right) =\alpha \left( {\bf r},t\right) +%
{\cal O}\left( \varepsilon ^{5}\right) $ and hence the considered forcing is
almost periodic of fundamental frequency $\omega _{0}$.

Under these conditions a multiple scale analysis is possible and we look for
asymptotic solutions to Eq. (\ref{modelu}) in the form 
\begin{equation}
{\bf u}\left( {\bf r},t\right) =\sum\nolimits_{m=1}^{\infty }\varepsilon ^{m}%
{\bf u}_{m}\left( X,Y,T\right) .  \label{expansion}
\end{equation}
We finally introduce Eqs. (\ref{mu}) and (\ref{alpha}--\ref{expansion}) into
Eq. (\ref{modelu}) making use of the following chain rules for
differentiation 
\begin{eqnarray*}
\partial _{t}{\bf u} &=&\sum\nolimits_{m=1}^{\infty }\varepsilon ^{m}\left(
\partial _{t}{\bf u}_{m}+\varepsilon ^{2}\partial _{T}{\bf u}_{m}\right) , \\
\nabla ^{2}{\bf u} &=&\sum\nolimits_{m=1}^{\infty }\varepsilon ^{m+2}\left(
\partial _{X}^{2}+\partial _{Y}^{2}\right) {\bf u}_{m},
\end{eqnarray*}
and solve at increasing orders in $\varepsilon $.

\section{The complex Ginzburg--Landau equation}

The general form of Eq. (\ref{modelu}) at any order $\varepsilon ^{m}$ is
found to be 
\begin{equation}
{\frak J}\left( {\bf u}_{m}\right) ={\bf g}_{m}\left( X,Y,T,t\right) ,
\label{orderm}
\end{equation}
where
\begin{equation}
{\frak J}\left( {\bf u}\right) \equiv \partial _{t}{\bf u}-{\cal J}_{0}\cdot 
{\bf u},
\end{equation}
and ${\bf g}_{m}$ does not depend on ${\bf u}_{m}$ (but on ${\bf u}_{n<m}$).
Clearly, as ${\frak J}\left( \exp \left( i\omega _{0}t\right) {\bf h}\right)
={\frak J}\left( \exp \left( -i\omega _{0}t\right) \overline{{\bf h}}\right)
=0$, see Eq. (\ref{r}), the solvability of Eq. (\ref{orderm}) requires 
\begin{equation}
\int_{t}^{t+2\pi /\omega _{0}}dt^{\prime }{\bf h}^{\dagger }\cdot {\bf g}%
_{m}\left( X,Y,T,t^{\prime }\right) \exp \left( -i\omega _{0}t^{\prime
}\right) =0,  \label{solvabilitym}
\end{equation}
(or its equivalent complex-conjugate) which ensures that ${\bf g}_{m}$ does
not contain secular terms (proportional to $\exp \left( i\omega _{0}t\right) 
{\bf h}$ or to $\exp \left( -i\omega _{0}t\right) \overline{{\bf h}}$). Once
condition (\ref{solvabilitym}) is verified, the asymptotic solution to Eq. (%
\ref{orderm}) reads 
\begin{eqnarray}
{\bf u}_{m}\left( X,Y,T,t\right)  &=&u_{m}\left( X,Y,T\right) \exp \left(
i\omega _{0}t\right) {\bf h}+\overline{u}_{m}\left( X,Y,T\right) \exp \left(
-i\omega _{0}t\right) \overline{{\bf h}}  \nonumber \\
&&+{\bf u}_{m}^{\bot }\left( X,Y,T,t\right) ,  \label{um}
\end{eqnarray}
where $u_{m}\left( X,Y,T\right) $ is a function of the slow scales, and the
last term is the particular solution. Note that the solution (\ref{um})
should involve, in principle, terms proportional to all the eigenvectors of $%
{\frak J}\left( \cdot \right) $ [which are those of ${\cal J}_{0}$, see Eq. (%
\ref{eigensystem})]. However all of them are damped according to $\exp \left[
-\left| 
\mathop{\rm Re}%
\Lambda _{i}\left( \mu _{0},0\right) \right| \,t\right] $, since $%
\mathop{\rm Re}%
\Lambda _{i\geq 3}\left( \mu =\mu _{0},k=0\right) <0$ by hypothesis, except
those associated with $\left( {\bf h},\overline{{\bf h}}\right) $.

\subsection{Order $\protect\varepsilon $}

This is the first nontrivial order and reads 
\begin{equation}
{\bf g}_{1}=0.  \label{order1}
\end{equation}
The solvability condition (\ref{solvabilitym}) at this order is fulfilled
and, according to Eq. (\ref{um}), 
\begin{equation}
{\bf u}_{1}=u_{1}\left( X,Y,T\right) \exp \left( i\omega _{0}t\right) {\bf h}%
+\overline{u}_{1}\left( X,Y,T\right) \exp \left( -i\omega _{0}t\right) 
\overline{{\bf h}}.  \label{u1}
\end{equation}

\subsection{Order $\protect\varepsilon ^{2}$}

At this order 
\begin{equation}
{\bf g}_{2}=\left( \partial _{\mu }{\bf F}\right) _{0}\mu _{2}+{\bf K}\left(
\mu _{0},0;{\bf u}_{1},{\bf u}_{1}\right) .  \label{order2}
\end{equation}
The first term of the r.h.s. is null by virtue of Eq. (\ref{Fmu0}). Making
use of Eq. (\ref{u1}) and taking into account that ${\bf K}$\ is symmetric
and bilinear in its two last arguments, Eq. (\ref{order2}) can be written as 
\begin{eqnarray}
{\bf g}_{2} &=&2{\bf K}\left( \mu _{0},0;{\bf h},\overline{{\bf h}}\right)
\left| u_{1}\right| ^{2}+{\bf K}\left( \mu _{0},0;{\bf h},{\bf h}\right)
u_{1}^{2}\exp \left( i2\omega _{0}t\right)   \nonumber \\
&&+{\bf K}\left( \mu _{0},0;\overline{{\bf h}},\overline{{\bf h}}\right) 
\overline{u}_{1}^{2}\exp \left( -i2\omega _{0}t\right) .
\end{eqnarray}
The solvability condition (\ref{solvabilitym}) is automatically fulfilled
again and, according to Eq. (\ref{um}), 
\begin{eqnarray}
{\bf u}_{2} &=&u_{2}\left( X,Y,T\right) \exp \left( i\omega _{0}t\right) 
{\bf h}+\overline{u}_{2}\left( X,Y,T\right) \exp \left( -i\omega
_{0}t\right) \overline{{\bf h}}  \nonumber \\
&&+{\bf v}_{0}\left| u_{1}\right| ^{2}+{\bf v}_{2}u_{1}^{2}\exp \left(
2i\omega _{0}t\right) +\overline{{\bf v}}_{2}\overline{u}_{1}^{2}\exp \left(
-2i\omega _{0}t\right) ,  \label{u2}
\end{eqnarray}
where, 
\begin{eqnarray}
{\bf v}_{0} &=&-2{\cal J}_{0}^{-1}\cdot {\bf K}\left( \mu _{0},0;{\bf h},%
\overline{{\bf h}}\right) ,  \label{v0} \\
{\bf v}_{2} &=&-\left( {\cal J}_{0}-i2\omega _{0}{\cal I}\right) ^{-1}\cdot 
{\bf K}\left( \mu _{0},0;{\bf h},{\bf h}\right) ,  \label{v2}
\end{eqnarray}
are constant vectors, and ${\cal I}$ is the $N\times N$ identity matrix.
Note that both ${\cal J}_{0}$ and $\left( {\cal J}_{0}-i2\omega _{0}{\cal I}%
\right) $ are invertible since neither $0$ nor $2i\omega _{0}$ are
eigenvalues of ${\cal J}_{0}$ by hypothesis: otherwise other eigenvalues
different from $\left\{ i\omega _{0},-i\omega _{0}\right\} $ would have null
real part at the bifurcation.

\subsection{Order $\protect\varepsilon ^{3}$}

Finally, at this order we find 
\begin{eqnarray}
{\bf g}_{3} &=&-\partial _{T}{\bf u}_{1}+\mu _{2}\left( \partial _{\mu }%
{\cal J}\right) _{0}\cdot {\bf u}_{1}+{\cal D}_{0}\cdot \left( \partial
_{X}^{2}+\partial _{Y}^{2}\right) {\bf u}_{1}  \nonumber \\
&&+\left( \partial _{\alpha }{\bf F}\right) _{0}\alpha _{3}+2{\bf K}\left(
\mu _{0},0;{\bf u}_{1},{\bf u}_{2}\right) +{\bf L}\left( \mu _{0},0;{\bf u}%
_{1},{\bf u}_{1},{\bf u}_{1}\right) .  \label{order3}
\end{eqnarray}
Application of the solvability condition (\ref{solvabilitym}) yields, after
substituting Eqs. (\ref{alpha3}), (\ref{u1}) and (\ref{u2}) into Eq. (\ref
{order3}), and making use of the symmetry and linearity properties of
vectors ${\bf K}$ and ${\bf L}$, 
\begin{equation}
\partial _{T}u_{1}=c_{1}\mu _{2}u_{1}+c_{2}\left( \partial _{X}^{2}+\partial
_{Y}^{2}\right) u_{1}+c_{3}\left| u_{1}\right| ^{2}u_{1}+c_{4}\alpha _{3,1},
\label{cgl1}
\end{equation}
where 
\begin{eqnarray}
c_{1} &=&{\bf h}^{\dagger }\cdot \left( \partial _{\mu }{\cal J}\right)
_{0}\cdot {\bf h},  \label{c1} \\
c_{2} &=&{\bf h}^{\dagger }\cdot {\cal D}_{0}\cdot {\bf h},  \label{c2} \\
c_{3} &=&2{\bf h}^{\dagger }\cdot {\bf K}\left( \mu _{0},0;{\bf h},{\bf v}%
_{0}\right) +2{\bf h^{\dagger }\cdot K}\left( \mu _{0},0;\overline{{\bf h}},%
{\bf v}_{2}\right) +3{\bf h}^{\dagger }\cdot {\bf L}\left( \mu _{0},0;{\bf h}%
,{\bf h},\overline{{\bf h}}\right) ,  \label{c3} \\
c_{4} &=&{\bf h}^{\dagger }\cdot \left( \partial _{\alpha }{\bf F}\right)
_{0},  \label{c4}
\end{eqnarray}
are constant coefficients.

Finally note that, making use of Eqs. (\ref{expansion}), (\ref{scales}) and (%
\ref{u1}), the asymptotic state of the system can be written as 
\begin{eqnarray}
{\bf u}\left( {\bf r},t\right)  &=&u\left( {\bf r},t\right) \exp \left(
i\omega _{0}t\right) {\bf h}+\overline{u}\left( {\bf r},t\right) \exp \left(
-i\omega _{0}t\right) \overline{{\bf h}}+{\cal O}\left( \varepsilon
^{2}\right) ,  \label{center} \\
u\left( {\bf r},t\right)  &=&\varepsilon u_{1}\left( X,Y,T,t\right) , 
\nonumber
\end{eqnarray}
hence $u$ denotes the leading order amplitude of oscillations. Its evolution
equation is obtained from Eq. (\ref{cgl1}) by returning to original scales
and parameters via Eqs. (\ref{mu}) and (\ref{scales}), and reads 
\begin{equation}
\partial _{t}u\left( {\bf r},t\right) =\left( \mu -\mu _{0}\right)
c_{1}u+c_{2}\nabla ^{2}u+c_{3}\left| u\right| ^{2}u+f\left( {\bf r},t\right)
,  \label{cgl}
\end{equation}
where 
\begin{equation}
f\left( {\bf r},t\right) =c_{4}\varepsilon ^{3}\alpha _{3,1}\left(
X,Y,T\right)   \label{f}
\end{equation}
is a slowly varying function proportional to the complex amplitude of the
fundamental component of forcing $\varepsilon ^{3}\alpha _{3,1}$, see Eqs. (%
\ref{alpha},\ref{alpha3}).

Eq. (\ref{cgl}), or (\ref{cgl1}), is a complex Ginzburg--Landau equation
containing an inhomogeneous forcing term, $f\left( {\bf r},t\right) $, which
generalizes Eq. (\ref{cglcoullet}) for $n=m=1$. Eq. (\ref{cgl}) is valid
whenever $%
\mathop{\rm Re}%
c_{3}\leq 0$ (supercritical bifurcation) since otherwise it could lead to
unbounded solutions. If $%
\mathop{\rm Re}%
c_{3}>0$ the bifurcation is subcritical and the analysis must incorporate
higher orders in the $\varepsilon $--expansion.

\section{Application to two-frequency forcing}

Just for the sake of illustration let us finally consider the two-frequency
forcing case. In particular we assume  that the forcing parameter $\alpha $
has the form
\begin{equation}
\alpha \left( {\bf r},t\right) =A\left( \omega _{1}t\right) +A\left( \omega
_{2}t\right) ,  \label{alpha12}
\end{equation}
where $A$ is a $2\pi $ periodic function $\left[ A\left( \theta \right)
=A\left( \theta +2\pi \right) \right] $. This means that we are dealing with
the superposition of two forcings of equal amplitude but of different
frequencies ($\omega _{1}$ and $\omega _{2}$). An example of this type of
forcing could be the illumination of a photosensitive version of the
Belousov-Zhabotinsky reaction with two equal sources, whose light
intensities are periodically modulated in time at two different frequencies.
Another example could be the injection into a laser cavity of two coherent
fields of equal amplitudes and different frequencies.

Due to the commented periodicity one can write
\begin{equation}
A\left( \omega _{i}t\right) =\sum\nolimits_{p=1}^{\infty }\left[ A_{p}\exp
\left( ip\omega _{i}t\right) +\bar{A}_{p}\exp \left( -ip\omega _{i}t\right) %
\right] .  \label{A}
\end{equation}
(Note that the term $p=0$ is absent, as in Eqs. (\ref{alpha},\ref{alpha3}),
since it would correspond to a constant bias. This bias term, if any, is
implicitly considered in our analysis at the same level as any other
constant parameter of the system and it would appear, in principle, in the
determination of the Hopf bifurcation). Upon rewriting Eq. (\ref{A}) as
\begin{equation}
A\left( \omega _{i}t\right) =\sum\nolimits_{p=1}^{\infty }\left[ A_{p}\exp
\left( ip\delta _{i}t\right) \exp \left( ip\omega _{0}t\right) +\bar{A}%
_{p}\exp \left( -ip\delta _{i}t\right) \exp \left( -ip\omega _{0}t\right) %
\right] ,  \label{A0}
\end{equation}
where $\delta _{i}=\omega _{i}-\omega _{0}$, we can express Eq. (\ref
{alpha12}) in the form (\ref{alpha},\ref{alpha3}): 
\begin{equation}
\alpha \left( {\bf r},t\right) =\sum\nolimits_{p=1}^{\infty }\left[
a_{p}\left( t\right) \exp \left( ip\omega _{0}t\right) +\bar{a}_{p}\left(
t\right) \exp \left( -ip\omega _{0}t\right) \right] ,  \label{alpha12b}
\end{equation}
where
\begin{equation}
a_{p}\left( t\right) =A_{p}\left[ \exp \left( ip\delta _{1}t\right) +\exp
\left( ip\delta _{2}t\right) \right] =2A_{p}\exp \left( ip\nu t\right) \cos
\left( p\omega t\right) ,  \label{ap}
\end{equation}
where
\begin{equation}
\nu \equiv \frac{\delta _{1}+\delta _{2}}{2}=\frac{\omega _{1}+\omega _{2}}{2%
}-\omega _{0},\quad \omega \equiv \frac{\delta _{1}-\delta _{2}}{2}=\frac{%
\omega _{1}-\omega _{2}}{2}.
\end{equation}
(We note incidentally that, in the case of single-frequency forcing, $\omega
_{1}=\omega _{2}\equiv \omega _{e}$ and hence $\nu =\omega _{e}-\omega _{0}$
($\omega _{e}=\omega _{0}+\nu $), and $\omega =0$.) Hence all our previous
analysis is valid for this type of driving, whenever forcing is weak ($%
A_{p}\sim \varepsilon ^{3}$) and the time scale along which the $a_{p}$'s
vary is slow. Specifically this requires $\nu ,\omega \sim \varepsilon ^{2}$%
. If these relations are satisfied, the system will be described by Eq. (\ref
{cgl}) with $f$ given by Eq. (\ref{f}) and $\varepsilon ^{3}\alpha _{3,1}$
given by $a_{p=1}\left( t\right) $:
\begin{equation}
\partial _{t}u\left( {\bf r},t\right) =\left( \mu -\mu _{0}\right)
c_{1}u+c_{2}\nabla ^{2}u+c_{3}\left| u\right| ^{2}u+2c_{4}A_{1}\exp \left(
i\nu t\right) \cos \left( \omega t\right) .  \label{cglcasi}
\end{equation}

In order to simplify this equation let us define
\begin{equation}
U\left( {\bf r},t\right) =u\left( {\bf r},t\right) \exp \left( -i\nu
t\right) ,  \label{Ugran}
\end{equation}
which, substituted into Eq. (\ref{cglcasi}) yields
\begin{equation}
\partial _{t}U\left( {\bf r},t\right) =\left[ \left( \mu -\mu _{0}\right)
c_{1}-i\nu \right] U+c_{2}\nabla ^{2}U+c_{3}\left| U\right|
^{2}U+2c_{4}A_{1}\cos \left( \omega t\right) .  \label{dtU}
\end{equation}
The actual state of the system, Eq. (\ref{center}), is given in terms of $U$
as
\begin{equation}
{\bf u}\left( {\bf r},t\right) =2%
\mathop{\rm Re}%
\left\{ U\left( {\bf r},t\right) \exp \left[ i\left( \omega _{0}+\nu \right)
t\right] {\bf h}\right\} .  \label{ufin1}
\end{equation}
Finally, note that, especially in nonlinear optics, one is used to express
the state of the system in terms of the negative-frequency part of the
oscillations, i.e., as
\begin{equation}
{\bf u}\left( {\bf r},t\right) =2%
\mathop{\rm Re}%
\left\{ \bar{U}\left( {\bf r},t\right) \exp \left[ -i\left( \omega _{0}+\nu
\right) t\right] \overline{{\bf h}}\right\} ,
\end{equation}
and the complex amplitude $\bar{U}\left( {\bf r},t\right) $ verifies the
complex conjugate of Eq. (\ref{dtU}):
\begin{equation}
\partial _{t}\bar{U}\left( {\bf r},t\right) =\left[ \left( \mu -\mu
_{0}\right) \bar{c}_{1}+i\nu \right] \bar{U}+\bar{c}_{2}\nabla ^{2}\bar{U}+%
\bar{c}_{3}\left| \bar{U}\right| ^{2}\bar{U}+2\bar{c}_{4}\bar{A}_{1}\cos
\left( \omega t\right) .
\end{equation}
This absolutely trivial transformation is done here just to point out that
the sign that affects the mistuning $\nu $ in the forced CGL equation
depends on whether one uses the amplitude of the positive- or the
negative-frequency part of the oscillations in order to describe the
dynamics of the system.

\end{document}